\documentclass[10pt,twocolumn,letterpaper]{article}

\usepackage{wacv}

\usepackage{amsmath}
\usepackage{amssymb}
\usepackage{booktabs}
\usepackage[english]{babel}
\usepackage[symbol, flushmargin]{footmisc}
\usepackage{epsfig}
\usepackage{graphicx}
\usepackage{multicol}
\usepackage{multirow}
\usepackage{listings}
\usepackage{mathabx}
\usepackage{subcaption}
\usepackage{times}
\usepackage{xcolor}

\usepackage{array}
\newcolumntype{L}[1]{>{\raggedright\let\newline\\\arraybackslash\hspace{0pt}}m{#1}}
\newcolumntype{C}[1]{>{\centering\let\newline\\\arraybackslash\hspace{0pt}}m{#1}}
\newcolumntype{R}[1]{>{\raggedleft\let\newline\\\arraybackslash\hspace{0pt}}m{#1}}

\usepackage[numbers,sort&compress]{natbib} % not sure we are allowed to use this

\wacvfinalcopy 

\ifwacvfinal
\usepackage[breaklinks=true,bookmarks=false]{hyperref}
\else
\usepackage[pagebackref=true,breaklinks=true,colorlinks,bookmarks=false]{hyperref}
\fi

\ifwacvfinal
\else
\fi

\def\etal{\emph{et al. }}

\renewcommand\v[1]{\mathbf{#1}}

\newcommand{\datadist}{p(x)}

\newcommand{\targets}{\v{x}}
\newcommand{\recons}{\xhat}
\newcommand{\latents}{\v{z}}
\newcommand{\target}[1]{\targets_{#1}}
\newcommand{\recon}[1]{\recons_{#1}}

\newcommand\flowpredicted[1]{\v{f}^{p}_{#1}}
\newcommand\flowdelta[1]{\deltafhat_{#1}}
\newcommand\totalflow[1]{\widehat{\v{f}}_{#1}}

\newcommand\respredicted[1]{\v{r}^{p}_{#1}}
\newcommand\resdelta[1]{\deltarhat_{#1}}

\newcommand\reconflowpredicted[1]{\target{#1}^p}
\newcommand\reconflow[1]{\target{#1}^w}
\newcommand\reconrespredicted[1]{\target{#1}^r}

\newcommand\xt{\v{x}_t}
\newcommand\xhat{\widehat{\v{x}}}
\newcommand\xhatt{\widehat{\v{x}}_{t}}
\newcommand\xhattmone{\widehat{\v{x}}_{t-1}}
\newcommand\xhattmtwo{\widehat{\v{x}}_{t-2}}
\newcommand\xpt{\v{x}^{p}_{t}}
\newcommand\xwt{\v{x}^{w}_{t}}

\newcommand\xwtmone{\v{x}^{w}_{t-1}}

\newcommand\fhatt{\widehat{\v{f}}_{t}}
\newcommand\deltafhat{\Delta\widehat{\v{f}}}
\newcommand\deltafhatt{\Delta\widehat{\v{f}}_{t}}
\newcommand\fpt{\v{f}^{p}_{t}}

\newcommand\rt{\v{r}_t}

\newcommand\deltarhat{\Delta\widehat{\v{r}}}
\newcommand\deltarhatt{\Delta\widehat{\v{r}}_{t}}
\newcommand\rpt{\v{r}^{p}_{t}}

\newcommand{\iframeae}{\operatorname{Image-AE}}
\newcommand{\flowae}{\operatorname{Flow-AE}}
\newcommand{\resae}{\operatorname{Res-AE}}
\newcommand{\flownet}{\operatorname{Flow-Pred}}
\newcommand{\resnet}{\operatorname{Res-Pred}}

\newcommand{\warp}{\operatorname{Warp}}

\newcommand{\mse}{\operatorname{MSE}}
\newcommand{\psnr}{\operatorname{PSNR}}
\newcommand{\msssim}{\operatorname{MS-SSIM}}

\begin{document}

%%%%%%%%% TITLE
\title{Boosting neural video codecs by exploiting hierarchical redundancy}

\author{Reza Pourreza, Hoang Le, Amir Said, Guillaume Sauti\`{e}re, Auke Wiggers \\
Qualcomm AI Research\footnotemark[3] \\
{\tt\small 
\{pourreza, hoanle, asaid, gsautie, auke\}@qti.qualcomm.org}
}

\maketitle
%\thispagestyle{empty}

%%%%%%%%% ABSTRACT
\begin{abstract}
In video compression, coding efficiency is improved by reusing pixels from previously decoded frames via motion and residual compensation. 
We define two levels of hierarchical redundancy in video frames: 
1) first-order: redundancy in pixel space, \ie, similarities in pixel values across neighboring frames, which is effectively captured using motion and residual compensation, 
2) second-order: redundancy in motion and residual maps due to smooth motion in natural videos. 
While most of the existing neural video coding literature addresses first-order redundancy, we tackle the problem of capturing second-order redundancy in neural video codecs via predictors. 
We introduce generic motion and residual predictors that learn to extrapolate from previously decoded data. 
These predictors are lightweight, and can be employed with most neural video codecs in order to improve their rate-distortion performance.
Moreover, while RGB is the dominant colorspace in neural video coding literature, we introduce general modifications for neural video codecs to embrace the YUV420 colorspace and report YUV420 results. 
Our experiments show that using our predictors with a well-known neural video codec leads to 38\% and 34\% bitrate savings in RGB and YUV420 colorspaces measured on the UVG dataset. 

\end{abstract}

\footnotetext[3]{
  Qualcomm AI Research is an initiative of Qualcomm Technologies, Inc.
}

% %%%%%%%%% BODY TEXT
\section{Introduction}

Video traffic is predicted to reach more than 82\% of all internet traffic in 2022 \cite{cisco2017visualnetworking}.
As video resolution and framerate demands have steadily increased over the years, there is a strong need to develop algorithms that enable transmission of video data at low bit cost.
Standard video codecs such as AVC \cite{wiegand2003overview} and HEVC \cite{sullivan2012hevc} were created exactly for this purpose.

The goal of a codec is to trade off the number of bits spent to transmit a message and the distortion between the original data and its reconstruction. 
In the video setting, standard codecs keep the bitrate low by using previously transmitted information as context. 
For example, to make a prediction for a current frame, the previously transmitted frame can be used.
Instead of transmitting the current frame, the motion between the previous reconstruction and the current frame may be transmitted.
This allows making an initial prediction for the current frame using motion compensation.
Transmitting the motion typically has a lower bit cost than transmitting the frame, as motion of pixels belonging to the same object are strongly correlated.
Similarly, an additive correction or \emph{residual} can be transmitted to refine a receiver-side frame. 
If the residual is sparse, the cost of transmitting a residual will be much lower than transmitting the frame in isolation.

Standard codecs take this idea one step further and make \emph{motion vector predictions}.
As motion vectors are typically similar across timesteps due to smooth object motion in video, an initial prediction can be made based on previous motion.
This means the codec only needs to transmit corrections to this prediction, usually at low bit cost. 

\begin{figure}[t!]
   \centering
    \includegraphics[width=0.83\linewidth]{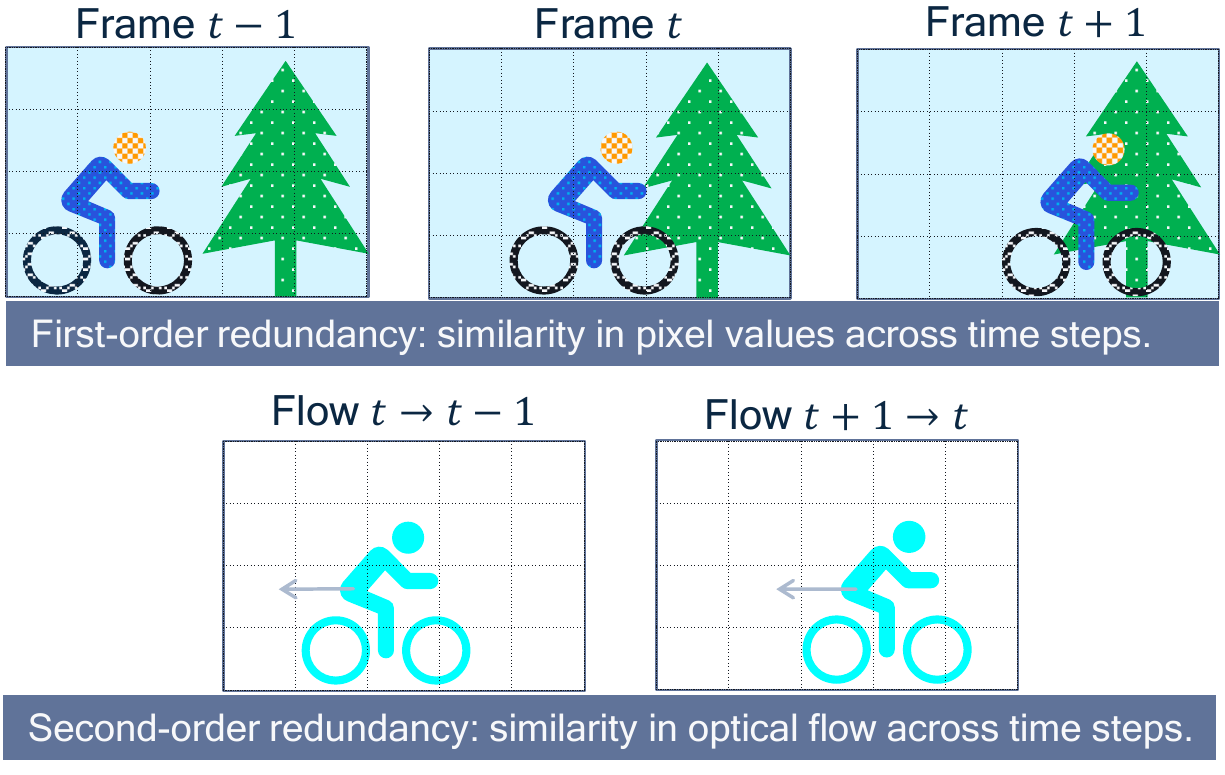}
    \vspace{-0.8em}
    \caption{
        First- and second-order redundancy in video, here shown for (backward) optical flow. First-order redundancy exists between frames in pixel-space, second-order redundancy exists between optical flow maps.
    }
    \label{fig:intro:redundancy}
    \vspace{-1.6em}
\end{figure}

To distinguish these two approaches, we distinguish between two types of redundancy in videos.
First-order redundancy is redundancy in the pixel domain.
Inter-frame codecs exploit this using motion compensation and residual coding.
Second-order redundancy is the redundancy in flow and residual maps due to smooth motions in videos, which can be exploited by making motion vector predictions. 
These are shown schematically for motion in Fig.~\ref{fig:intro:redundancy}.

Most modern neural video codecs exploit only first-order redundancy using a two-step process \cite{dvc,ssf,hu2021fvc,hu2022coarse}:
a \emph{flow model} transmits motion information (optical flow) and \emph{residual model} transmits residual maps.
There are works that exploit second-order redundancy as well.
For example, ELF-VC~\cite{elfvc} predicts the flow, and
Liu \etal \cite{liu2021residualprediction} use flow and residual prediction in latent space.

In this work, we exploit second-order redundancy via pixel-space flow and residual predictors.
We design generic models that predict flow and residual using the history available in a frame buffer. 
This leads to easy-to-interpret predictions compared to predictions in latent space, makes it possible to combine the predictors with many existing neural codecs, and delivers significant performance improvement.

Furthermore, we provide evaluations for multiple input colorspaces. 
Most literature on neural video codecs only provides solutions and evaluations for the RGB colorspace, with some exceptions \cite{egilmez2021, singh2021combined}.
However, standard codecs were designed to operate with YUV420 input. 
YUV420 is a color format that subsamples the chroma channels of the YUV colorspace which is closer to how humans perceive images and leads to bandwidth saving in compression applications.
To gauge progress in learned video compression compared to standard codecs, we train our codecs on both RGB and YUV420 input formats.
We demonstrate that neural codecs can excel in both domains, and that only minor architectural modifications are needed to handle the difference in input resolution between colorspaces.

In summary, the contributions of this paper are:
\begin{enumerate}
    \itemsep0em 
    \item We define the framework of hierarchical redundancy in video in order to categorize neural codecs, and observe that most neural codecs only address the first type of redundancy.
    \item To exploit second-order redundancy, we introduce generic flow and residual predictors that can be paired with existing inter-frame codecs and deliver strong performance improvements. 
    \item We show that simple modifications enable codecs to run on YUV420 inputs, and report both RGB and YUV420 performance.
\end{enumerate}

%%%%%%%%%%%%%%%%%%%%%%%%%%%%%%%%%%%%%%%%%%%%%%%%%%%%%%%%%%%%%%%%%%%%%%%%%%%%%%%%
\section{Related work}

%-------------
\subsection{Learned compression}

The focus of this work is a so-called \emph{learned codec}, meaning each component is a learned model.
Neural network-based codecs have been successfully applied to data compression in many domains, including the image \cite{toderici2017full,theis2017lossy,agustsson2019extreme} and video \cite{dvc,habibian2019video,ssf,hu2021fvc,elfvc,hu2022coarse} domain.
In these systems, a neural encoder takes the data $\targets$ and produces a quantized latent variable $\latents$, and a neural decoder produces a reconstruction $\recons$ given this latent.
A neural prior or context model is used to learn the distribution of latent variables $p(z)$.
Using this prior and an entropy coding algorithm, Shannons source-coding theory tells us that the latent $\latents$ can be losslessly compressed using $-\log p(\latents)$ bits.

Neural codecs are typically trained to minimize a rate-distortion loss consisting of two terms:
\begin{equation}
    \setlength\abovedisplayskip{3pt}
    \mathcal{L}_{RD} = \mathbb{E}_{\targets \sim \datadist} \left[ \beta \mathcal{L}_{rate}( \latents ) + \mathcal{L}_{distortion}(\targets, \recons) \right].
    \setlength\belowdisplayskip{3pt}
\end{equation}

The rate term corresponds to the number of bits needed to transmit the quantized latent variable $\latents$, and a distortion term corresponds to the distance between the reconstruction $\recons$ and the ground truth $\targets$.

Neural codecs promise several advantages over handcrafted codecs.
First, as they learn to identify redundancies from example data, they have shown a strong ability to specialize to a domain \cite{habibian2019video} or even a single datapoint~\cite{rozendaal2021overfitting,strumpler2021implicit,rozendaal2021instance}. 
For example, if the codec will only be used to code animated content, it can easily be finetuned to this domain. 
Second, neural codecs benefit from advances in general-purpose neural hardware, where most video codecs require dedicated hardware to run in real time on device.
Third, they have shown to be able to hallucinate desirable textures, leading to improved perceptual quality compared to traditional codecs in user studies
\cite{mentzer2020hific,mentzer2021generative,yang2021perceptual}. 

%-------------
\subsection{Neural video compression}

In the video setting, neural codec design has been inspired by techniques from handcrafted codecs.
Where early works used frame interpolation \cite{wu2018video} or predicted entire blocks of frames jointly \cite{habibian2019video}, more recent neural video codecs exploit similarity between frames using motion compensation and residual coding.
Having neural networks learn to perform these steps end-to-end, as introduced by \cite{dvc}, has led to major bitrate savings in both the low latency \cite{ssf,singh2021combined,hu2021fvc,elfvc,hu2022coarse}, and streaming setting \cite{pourreza2021extending,ladune2021conditional}.

As consecutive frames often contain continuous motion, both motion information and residuals are predictable given previous frames.
Recent neural video codecs exploit this fact, for example, by adding components that predict the flow \cite{elfvc} or predict the motion and residual latents \cite{liu2021residualprediction}.

In this work, we use simple predictors for both flow and residual.
By only incorporating context from the past two timesteps, and avoiding statefulness and recurrent components, we avoid instabilities due to aggregating error.
Unlike ELF-VC \cite{elfvc}, our codec requires no normalization layers anywhere in the network, which makes the codec easier to deploy to hardware \cite{nagel2021white,thiruvathukal2022thiruvathukal}.
Unlike Liu \cite{liu2021residualprediction}, we make predictions in pixel-space. This makes predictions easy to interpret, and enables combining our predictors with different base models. 

%-------------
\subsection{YUV420 colorspace}
Most neural video codecs are trained to maximize PSNR or MS-SSIM in the RGB colorspace.
However, when visual quality according to human observers is important, other colorspaces may be better suited.
Distance in the YUV colorspace is known to match human perception more closely than distance in the euclidean RGB space \cite{tasic2003colorspaces}.

Most standard codecs were designed to maximize PSNR in the YUV420 input domain.
This is a subsampling of YUV, where chroma components are subsampled by $4\times$ to reduce bitrate and to meet bandwidth constraints.
For a $H \times W$ input, this subsampling results in one $H \times W$ luma channel, and two $ \frac{H}{2} \times \frac{W}{2}$ chroma channels.

There exist a few learned video codecs that support the YUV420 color format \cite{egilmez2021,singh2021combined}, or have been trained to optimize metrics in the YUV domain \cite{elfvc}.
However, as of yet, no learned video codec has shown competitive PSNR or MS-SSIM in the YUV420 domain.

\section{Method}

%---------------------
\subsection{Base codec}
\begin{figure}[t!]
    \centering
    \includegraphics[width=0.7\linewidth]{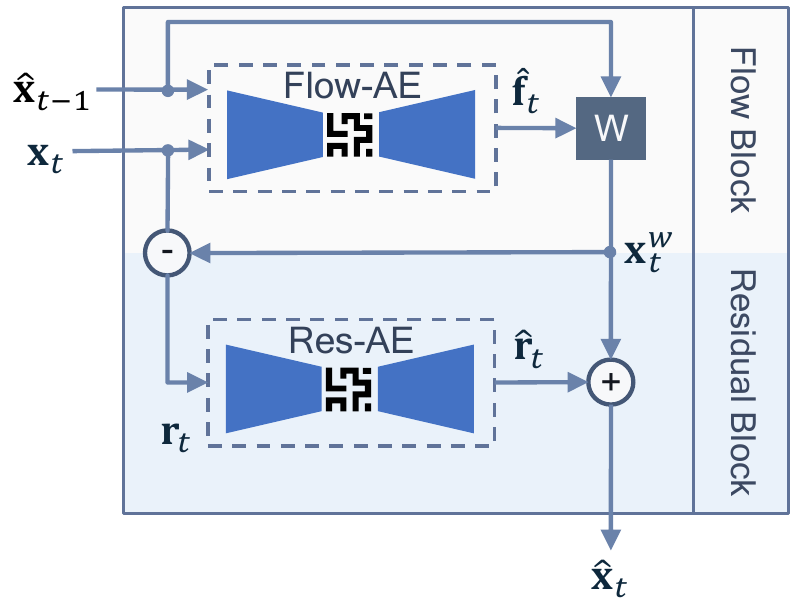}
    \vspace{-0.5em}
    \caption{Base inter-frame codec architecture performing motion compensation and residual coding. \texttt{W} indicates a scale-space warp.
    }
    \label{fig:introduction:base_codec}
    \vspace{-1.5em}
\end{figure}

In this work, we introduce flow and residual predictors to capture second-order redundancy in the low-delay neural video coding setting. 
Although the introduced predictors can potentially be added to any neural video codec, here, we build upon the well-known video codec scale-space flow (SSF)~\cite{ssf} due to the simple and efficient network design and the straightforward training schedule. 
The block-diagram of SSF in shown is Figure~\ref{fig:introduction:base_codec}, where $\fhatt$ indicates scale-space flow and $\warp$ indicates scale-space warping.

Figure~\ref{fig:introduction:redundancy} (a) shows two consecutive video frames, and the associated optical flow and residual maps are shown in Figure~\ref{fig:introduction:redundancy} (b) and (c), respectively. 
Of course the flow and residual for the first timestep depend on a previous frame, which is omitted here for brevity.
As can be seen from these figures, optical flow maps and residuals are much less detailed than video frames, making them easier to compress.
Additionally, due to the smooth motion in this video, optical flow and residual maps are highly correlated across the two time steps. 

%%%%%%%%%%%%%% Figure: base architecture

%%%%%%%%%%%%%% Figure: our architecture
\begin{figure}[t!]
    \centering
    \includegraphics[width=0.78\linewidth]{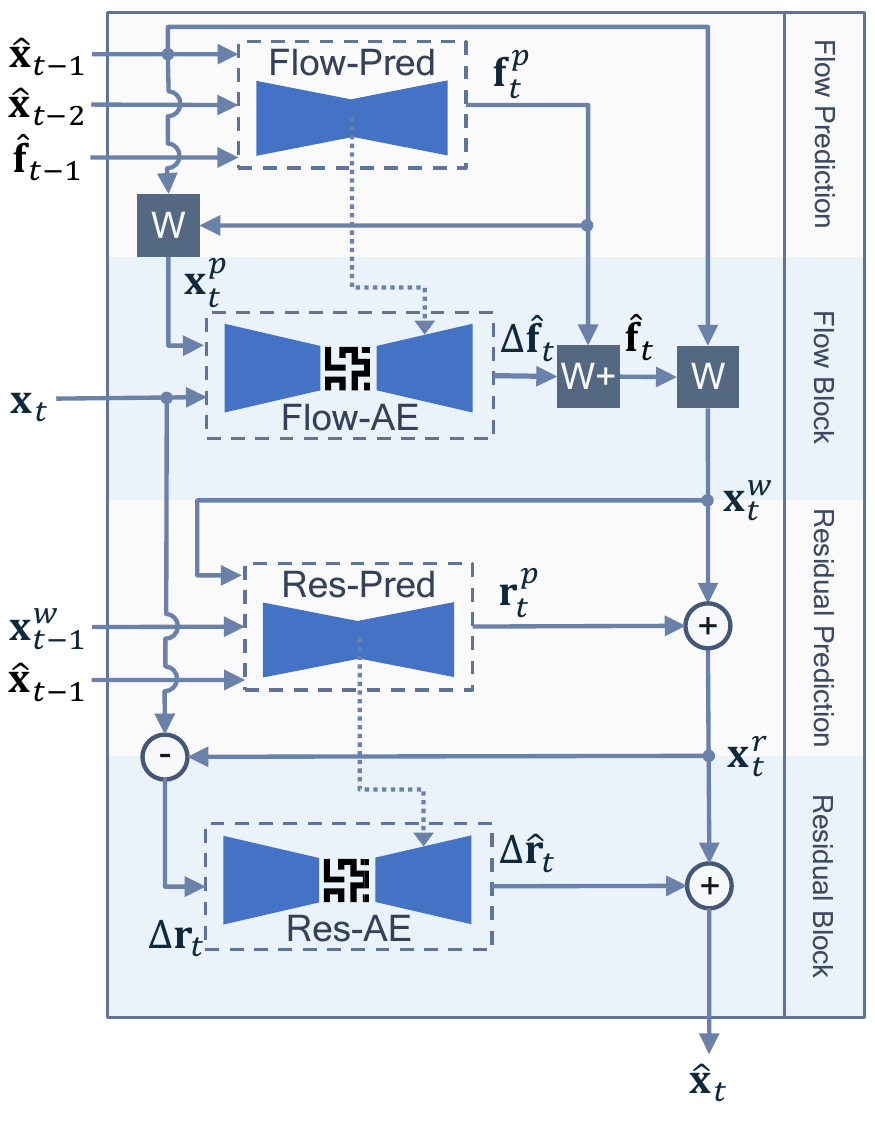}
    \vspace{-0.5em}
    \caption{Architecture for our inter-frame model.  \texttt{W+} indicates a scale-space warp followed by a summation. 
             }
    \label{fig:method:architecture:visualization}
    \vspace{-1.5em}
\end{figure}

%---------------------
\subsection{Predictors}
Here, we build a neural video codec on top of SSF where we keep the intra-frame codec as well the \emph{flow-block} and \emph{residual-block} in the inter-frame codec unchanged. 
We add flow and residual predictors to the base inter-frame codec as shown in Figure~\ref{fig:method:architecture:visualization}.

Inter-frame coding is done via a four-step process as follows: 1) \emph{flow prediction}, 2) \emph{flow compression}, 3) \emph{residual prediction}, 4) \emph{residual compression}. These four steps are explained below:

\begin{enumerate}
    \itemsep0em 
    \item Flow prediction: as the name suggests, we predict the flow for the current time step $\fpt$ using a history of decoded frames, $\xhattmone$ and $\xhattmtwo$, and the previous flow $\totalflow{t-1}$, as follows:
        \begin{equation}
          \setlength\abovedisplayskip{3pt}
          \label{eq:method:flowpred}
               \flowpredicted{t} = \flownet(\recon{t-2}, \recon{t-1},  \totalflow{t-1})
          \setlength\belowdisplayskip{3pt}
        \end{equation}
        Here, $\flownet$ is a lightweight network that directly predicts $\fpt$ using the provided inputs. The predicted flow $\fpt$ is then used to generate a prediction of the current frame $\xpt$. This operation is done only on the transmitter side.
        \begin{equation}
          \setlength\abovedisplayskip{3pt}
          \label{eq:method:flowpredapply}
               \reconflowpredicted{t} = \warp(\recon{t-1}, \flowpredicted{t})
          \setlength\belowdisplayskip{3pt}
        \end{equation}
    \item Flow compression: in this step, the predicted flow $\fpt$ is corrected using delta flow $\deltafhatt$ that is measured and transmitted via $\flowae$:
        \begin{align}
          \label{eq:method:flowblock}
            \setlength\abovedisplayskip{3pt}
            \flowdelta{t}          &= \flowae(\reconflowpredicted{t}, \xt) \\
            \fhatt                 &= \deltafhatt+\warp(\fpt, \deltafhatt)
            \setlength\belowdisplayskip{3pt}
        \end{align}
        Here, $\flowae$ is a hyperprior network identical to the one in SSF. The motion compensated frame $\xwt$ is generated using the corrected flow $\fhatt$:
        \begin{equation}
          \setlength\abovedisplayskip{3pt}
          \label{eq:method:flowblockapply}
            \reconflow{t} = \warp(\recon{t-1}, \fhatt)
          \setlength\belowdisplayskip{3pt}
        \end{equation}
    \item Residual prediction: we predict the residual for the current time step $\respredicted{t}$ based on the motion compensated frames $\xwt$ and $\xwtmone$ as well as the previous decoded frame $\xhattmone$:
        \begin{equation}
          \setlength\abovedisplayskip{3pt}
          \label{eq:method:respred}
            \respredicted{t} = \resnet(\reconflow{t}, \reconflow{t-1}, \recon{t-1}) 
          \setlength\belowdisplayskip{3pt}
        \end{equation}
        Here, $\resnet$ is a lightweight network that directly predicts $\rpt$ using the provided inputs. Given that $\rt=\xt-\xwt$, the rationale is that $\resnet$ should be able to predict $\rt$ by looking at $\xhattmone$ and $\xwtmone$. Next, $\rpt$ is applied to $\xwt$:
        \begin{equation}
          \setlength\abovedisplayskip{3pt}
          \label{eq:method:respredcomp}
            \reconrespredicted{t} = \reconflow{t} + \respredicted{t}
          \setlength\belowdisplayskip{3pt}
        \end{equation}
    \item Residual compression: in this step, the predicted residual $\rpt$ is corrected using delta residual $\deltarhatt$ that is transmitted via $\resae$::
        \begin{equation}
          \setlength\abovedisplayskip{3pt}
          \label{eq:method:resblock}
            \resdelta{t} = \resae(\xt - \reconrespredicted{t})
          \setlength\belowdisplayskip{3pt}
        \end{equation}
        Here, $\resae$ is a hyperprior network identical to the one in SSF. Finally, the decoded frame $\xhatt$ is generated as follows:
        \begin{equation}
          \setlength\abovedisplayskip{3pt}
          \label{eq:method:resblockapply}
            \recon{t} = \reconrespredicted{t} + \resdelta{t}
          \setlength\belowdisplayskip{3pt}
        \end{equation}
\end{enumerate}

\footnotetext[1]{
  \tiny{Video produced by Netflix, with \texttt{CC BY-NC-ND 4.0} license: \\
  \url{https://media.xiph.org/video/derf/ElFuente/Netflix_Tango_Copyright.txt}}
}

The following points are worth emphasizing:
\begin{itemize}
    \itemsep0em 
    \item All the used flows are scale-space flow and all the warp operations are scale-space warp \cite{ssf}.
    \item Both $\flownet$ and $\resnet$ have hourglass architectures. The decoders of $\flowae$ and $\resae$ are conditioned on the bottleneck features of $\flownet$ and $\resnet$, respectively.
    \item We always encode the first frame in a sequence as an intra-frame and the subsequent frames as inter-frames. For the first inter-frame $\target{1}$, note that $\recon{t-2}$, $\totalflow{t-1}$, and $\reconflow{t-1}$ are not available.
    They are set as follows:
    \begin{align}
      \setlength\abovedisplayskip{3pt}
      \label{eq:method:firstpframe}
        \recon{t-2}      &= \reconflow{t-1} = \recon{t-1} \\
        \totalflow{t-1}  &= \textbf{0}
      \setlength\belowdisplayskip{3pt}
    \end{align}
    \item To construct the total flow $\deltafhatt$ in step 2, we choose to warp the predicted flow, which we observed to lead to better performance than additive operations. Other schemes have been used as well, such as letting by the $\flowae$ output a scale-and-shift \cite{elfvc}. 
\end{itemize}

%%%%%%%%%%%%%% Figure: video frames, flow, residual
\begin{figure}[t!]
    \centering
    \includegraphics[width=0.99\linewidth]{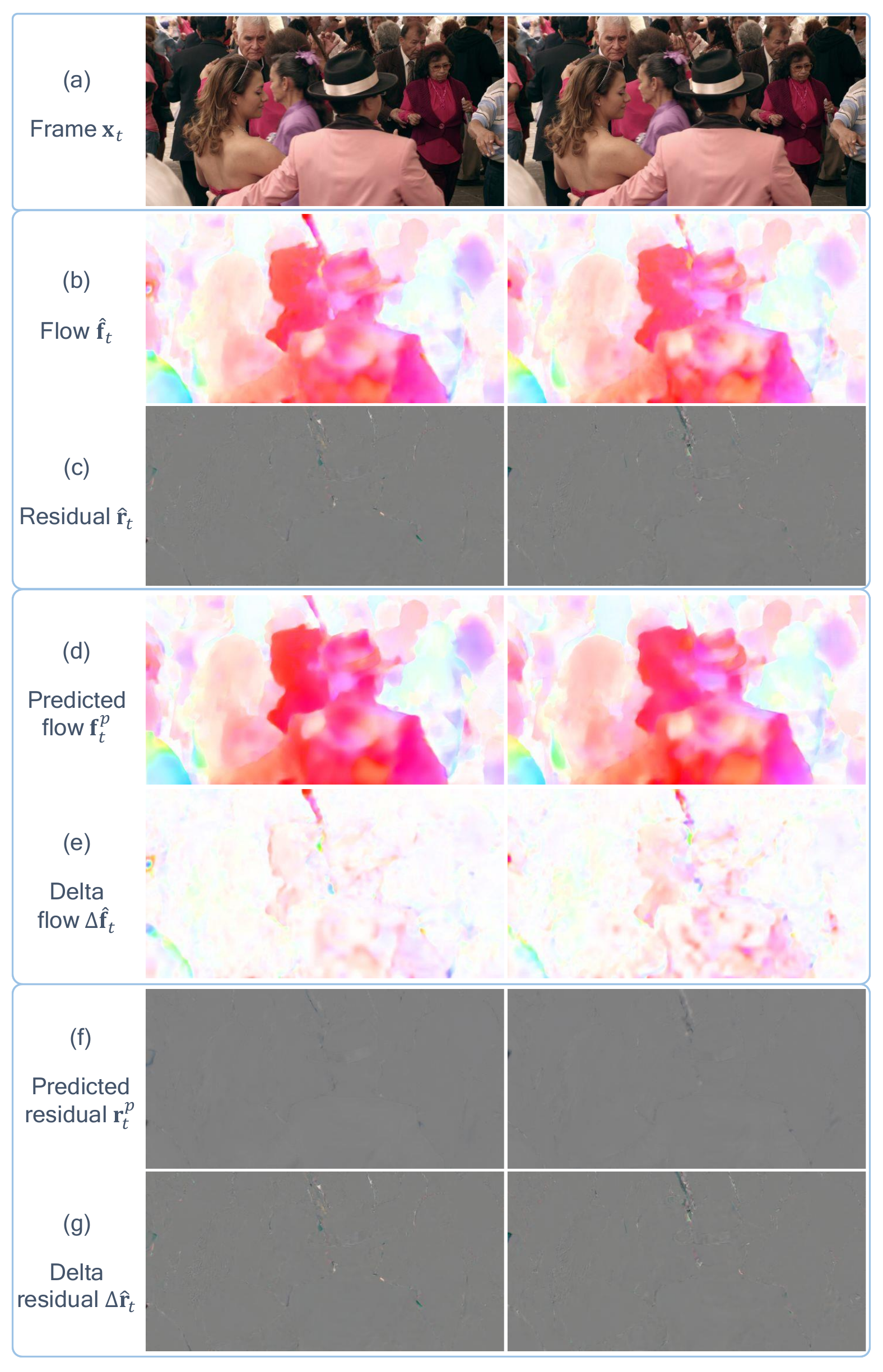}
    \vspace{-0.5em}
    \caption{
        Illustration of video (a), flow (b) and residual (c), flow prediction and delta flow (d, e), and residual prediction and delta residual (f, g). Screenshot from Tango video from Netflix Tango in Netflix El Fuente\protect\footnotemark[1]~\cite{Xiph}
    }
    \label{fig:introduction:redundancy}
    \vspace{-1.5em}
\end{figure}

Figure~\ref{fig:introduction:redundancy} (d) and (f) show the predicted flow and residual, respectively. 
The associated delta flow and delta residual are shown in Figure~\ref{fig:introduction:redundancy} (e) and (g). 
As can be seen here, the predicted flow and residual maps are very close to the actual maps, leaving much less correlated delta flow and delta residual.

\begin{figure}[t!]
    \centering
    \includegraphics[width=0.82\linewidth]{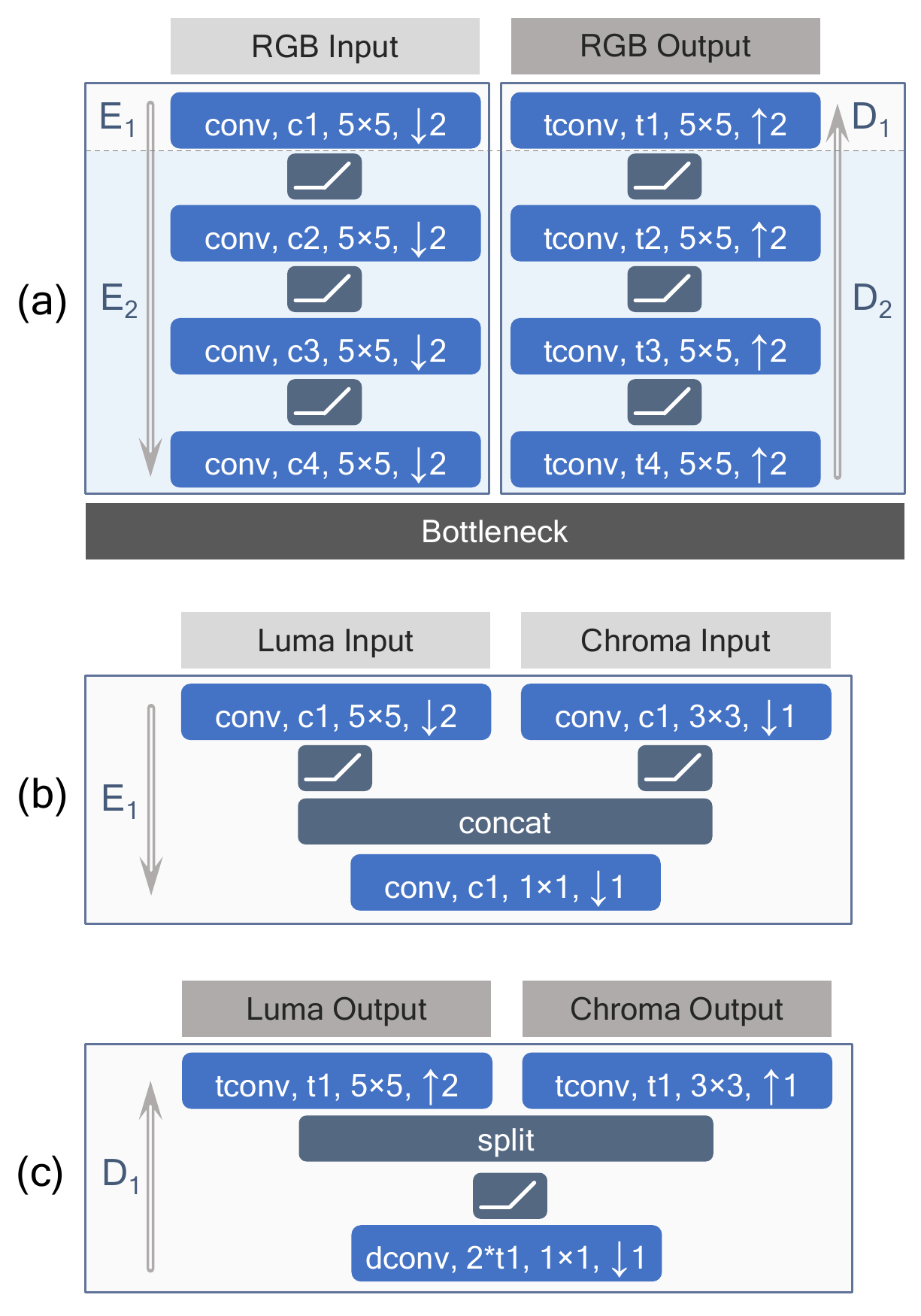}
    \vspace{-0.5em}
    \caption{a) Network architecture for the RGB colorspace, b) YUV420 input head, c) YUV420 output head. \texttt{c*} and \texttt{t*} indicate the number of filters in the convolutional and transposed-convolutional layers, respectively.
    }
    \label{fig:method:architecture:input_types}
    \vspace{-1.5em}
\end{figure}

%-----------------------------------------
\subsection{RGB and YUV420 architecture}

Most literature on neural video codecs provides solutions for the RGB colorspace. Since the R, G, and B channels share the same resolution, working with RGB videos is trivial as the networks can conveniently concatenate them in the network input and output. However, standard codecs were designed to operate with YUV420 input mainly due to the resemblance with human vision. Since the chroma channels in the YUV420 colorspace are in a subsampled space, some modifications to the network architecture are needed to accommodate YUV420 videos. Figure~\ref{fig:method:architecture:input_types} (a) shows the general architecture we used in the predictors and the autoencoders of our network for the RGB colorspace. The input is parsed by an encoder input head E1 and passed through an encoder backbone E2. The decoder consists of a backbone D2 and a decoder output head D1. At the bottleneck, the predictors simply pass the encoder outputs directly to the decoder. Components that transmit information will quantize the latents before passing these to the decoder.

To enable the codec with accommodating YUV420 inputs/outputs, we update the input/output heads E1/D1 in all the components of the network as shown in Figure ~\ref{fig:method:architecture:input_types} (b)/(c). E1 and D1 handle the luma (Y) and chroma (U and V) channels separately, and only the luma channels are down- or upsampled. Lastly, for the YUV420 codec, warping is applied for the luma and chroma channels separately. This ensures warping is always performed on the highest available spatial resolution.

\section{Experiments}

%-----------------------------------------
\subsection{Network architecture}
\label{sec:results:architecture}
All of the components in our codec follow the architecture shown in Figure~\ref{fig:method:architecture:input_types}.
The the number of filters in convolutional and transposed-convolutional layers is specified by \texttt{c1}, ..., \texttt{c4} and \texttt{t1}, ..., \texttt{t4}, respectively. 
More details are provided in Table~\ref{tab:experiments:compute:architecture}. 
In the main autoencoders -- including $\flowae$, $\resae$, and $\iframeae$ (used for intra-frame coding) -- the prior model follows the SSF design.

%%%%%%%%%%%%%%%%%%%%%%%%%%%%%%%%%%%%%%
\begin{table}[t!]
    \centering
    \small
\begin{tabular}{lccccc}
\toprule
  Module       & \texttt{c1} \& \texttt{t2} & \texttt{c2} \& \texttt{t3} & \texttt{c3} \& \texttt{t4} & \texttt{c4}  \\ 
\midrule
 $\iframeae$   &                128         &               128          &                   128      &       192    \\ 
 $\flowae$     &                128         &               128          &                   128      &       192    \\ 
 $\resae$      &                128         &               128          &                   128      &       192    \\ 
\midrule
 $\flownet$    &                 16         &               32           &                   64       &       128    \\ 
 $\resnet$     &                 16         &               32           &                   64       &       128    \\ 
\bottomrule
\end{tabular} 
    \vspace{-0.5em}
    \caption{Network architecture details. For all components, \texttt{t1} depends on the number of output channels.}
    \label{tab:experiments:compute:architecture}
\end{table}

%%%%%%%%%%%%%%%%%%%%%%%%%%%%%%%%%%%%%%
\begin{table}[t!]
    \centering
    \small
\begin{tabular}{clrrrr}
\toprule
 & Module           & \multicolumn{2}{l}{Parameters} & \multicolumn{2}{l}{KMACs / pixel} \\ 
\midrule
\parbox[t]{2mm}{\multirow{5}{*}{\rotatebox[origin=c]{90}{RGB}}}
& $\iframeae$    &   9.414M &    30.4\% &   198.2  &  30.6\% \\
& $\flowae$      &  10.038M &    32.4\% &   210.2  &  32.4\% \\
& $\resae$       &  10.028M &    32.5\% &   207.8  &  32.0\% \\
& $\flownet$     &   0.748M &     2.4\% &    16.2  &   2.5\% \\
& $\resnet$      &   0.748M &     2.4\% &    16.2  &   2.5\% \\
\midrule
\parbox[t]{2mm}{\multirow{5}{*}{\rotatebox[origin=c]{90}{YUV420}}}
& $\iframeae$    &    9.496M &   30.4\% &   208.0  &  30.5\% \\
& $\flowae$      &   10.124M &   32.4\% &   225.7  &  33.0\% \\
& $\resae$       &   10.111M &   32.3\% &   217.6  &  31.9\% \\
& $\flownet$     &    0.751M &    2.4\% &    16.3  &   2.4\% \\
& $\resnet$      &    0.750M &    2.4\% &    15.3  &   2.2\% \\
\bottomrule
\end{tabular} 
    \vspace{-0.5em}
    \caption{Parameters and MACs per pixel of our RGB (top) and YUV420 (bottom) codecs.}
    \label{tab:experiments:compute:complexity}
    \vspace{-1.5em}
\end{table} 

%------------
We report the computational complexity of our codec in Table~\ref{tab:experiments:compute:complexity}, where we show number of parameters and the number of multiply-and-accumulate operations (MACs) per module.
The MAC count is normalized by number of input pixels to enable comparison across input modalities and resolutions.
We also show the corresponding percentage to put these numbers in context.

We emphasize that for each I-frame, only the $\iframeae$ is used, and this model is not used for P-frames.
For example, for a sequence of length 10, the total number of MACs for the RGB codec would be $1 \cdot 198.2 + 9 \cdot ( 210.2 + 207.8 + 16.2 + 16.1 )$ KMACs per pixel.

Using MACs as a proxy for computational complexity, we observe that receiver-side predictors incur a small computational cost of only 2.5\% of the entire model.
Additionally, for each P-frame, both predictors incur $12\times$ fewer MACs than the P-frame autoencoders.
Although the YUV420 input has fewer dimensions than the RGB input due to the subsampled chroma components, the fact that the YUV420 codec processes luma and chroma channels using separate layers results in a higher parameter count and MAC count than that of the RGB codec.

%-----------------------------------------
\subsection{Dataset}
\label{sec:results:dataset}

Training is performed on the Vimeo90K dataset \cite{xue2019vimeodataset}, a diverse set of 89,800 video sequences of resolution $256 \times 448$ in RGB. 
We evaluate on three common video compression benchmark datasets:
UVG \cite{mercat20202uvg}, MCL-JCV \cite{wang2016mcljvc}, and
class B of HEVC (Common Test Conditions) \cite{bossen2013common}, all available in raw YUV420 format. 
UVG contains 7 full-HD (\ie, $1920\times1080$) videos at framerates of up to 120 fps.
MCL-JCV contains 30 full-HD videos at 30 fps.
HEVC class B contains 5 full-HD videos at various framerates.

%-----------------------------------------
\subsection{Training}

We train codecs with batch size of 8, where each sequences has $4$ frames of size $256 \times 256$, resulting in one I-frame and three P-frames.
We optimize our codec on $\mse$ and $\msssim$.
Specifically, we initially train all models for 1,000,000 gradient updates on $\mse$, then trained the $\msssim$ models for extra 200,000 gradient updates on $\msssim$, and finally finetuned all the models for 100,000 gradient updates.
Training is done on $256 \times 256$ patches with learning rate $10^{-4}$, and finetuning is done on $256 \times 384$ patches with learning rate $10^{-5}$, using the Adam optimizer.

For the RGB codec training, Video90k is used in the native RGB colorspace, and both $\mse$ and $\msssim$ are averaged over the R, G, and B channels.

For the YUV420 codec training, Video90k is converted to YUV420 using ffmpeg, and we weigh the $\mse$ and $\msssim$ for the luma component by a factor $\frac{6}{8}$ and the chroma components by a factor $\frac{1}{8}$, consistent with the evaluation procedure used for standard codecs~\cite{strom2020yuv420}. 

%-----------------------------------------
\subsection{Evaluation}

We evaluate all models in terms of rate-distortion performance. 
The $\psnr$ metric in RGB weighs all color-channels equally. 
For YUV420 input, the luma channel is weighed by a factor $\frac{6}{8}$, and both chroma channels by a factor $\frac{1}{8}$.
This is consistent with the way this metric is used in standard codecs~\cite{strom2020yuv420}.
As there is no commonly agreed-upon way to measure $\msssim$ for YUV420 inputs, we simply follow the same procedure as for $\psnr$, re-weighting the $\msssim$ for the luma and chroma components.

As the evaluation datasets are available in raw YUV420 format, they are used in the native colorspace for the YUV420 codec evaluation, see Figure~\ref{fig:experiments:results:evalsetup} (a). However, evaluating in RGB requires a conversion from YUV420 to RGB using ffmpeg as shown in Figure~\ref{fig:experiments:results:evalsetup} (b).

Most standard codecs were designed to work in the YUV420 input domain.
To enable a fair comparison to standard codecs, we always feed them YUV420 inputs and always operate them in the YUV420 colorspace. 
The YUV420 output is converted to the target colorspace (if needed) as shown in Figure~\ref{fig:experiments:results:evalsetup} (a) and (b) for comparison with our codecs in YUV420 and RGB, respectively.

\begin{figure}[t!]
    \centering
    \includegraphics[width=0.99\linewidth]{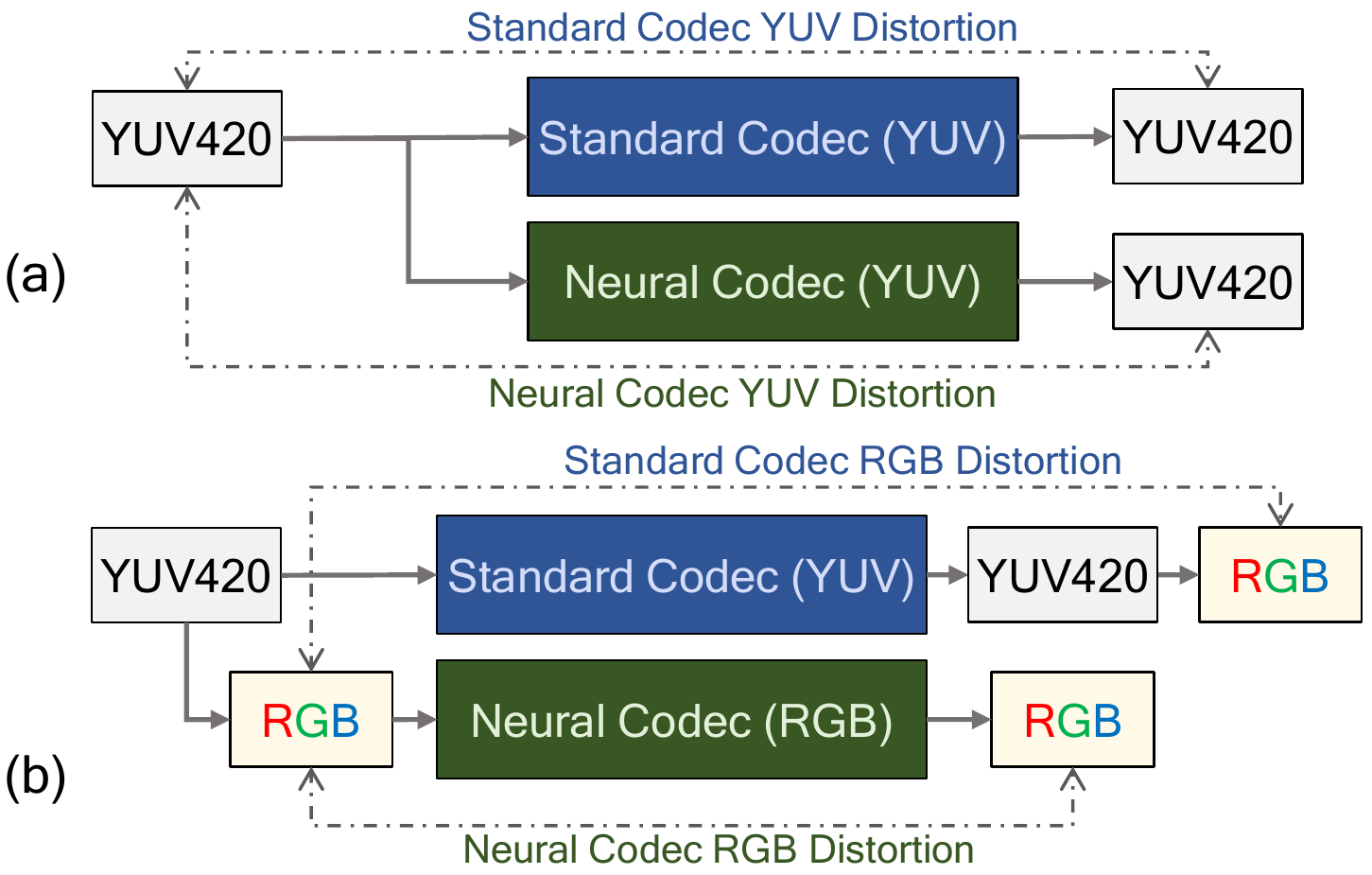}
    \vspace{-0.5em}
    \caption{Evaluation procedure for YUV420 (a) and RGB (b).
             Color-space conversions are performed using ffmpeg.
    }
    \label{fig:experiments:results:evalsetup}
    \vspace{-1.5em}
\end{figure}

%-----------------------------------------
\subsection{Results}

\begin{table}[b!]
  \centering 
  \small
    \begin{tabular}{l R{1.4cm} R{1.4cm} R{1.4cm}}
        \toprule
            & UVG       & HEVC-B    & MCL-JCV \\ 
        \midrule
    Average & -38.01\%  & -26.51\%  & -16.48\%  \\
        \bottomrule
    \end{tabular}
\vspace{-0.5em}
\caption{BD rate savings for RGB PSNR with with respect to the SSF base model. Lower is better.}
\label{tab:experiments:results:bdrates_rgb}
\end{table}

\begin{table}[b!]
  \centering
  \small
    \begin{tabular}{l R{1.4cm} R{1.4cm} R{1.4cm}}
        \toprule
                    & UVG       & HEVC-B    & MCL-JCV \\ 
        \midrule
        Y           & -28.34\%  & -28.50\%  & -21.85\%  \\
        U           & -47.32\%  & -44.38\%  & -34.63\%  \\
        V           & -55.68\%  & -43.72\%  & -36.51\%  \\
        \midrule
        Average     & -34.13\%  & -32.39\%  & -25.28\%  \\
        \bottomrule
    \end{tabular}
\vspace{-0.5em}
\caption{BD-rate savings for YUV420 PSNR with respect to the SSF-YUV model. 
For the average results, the luma and chroma BD-rates are weighed by $\frac{6}{8}$ and $\frac{1}{8}$ respectively. Lower is better. }
\label{tab:experiments:results:bdrates_yuv}
\end{table}

\begin{figure*}
    \begin{subfigure}{1.0\textwidth}
      \raggedleft
      \includegraphics[width=.98\textwidth]{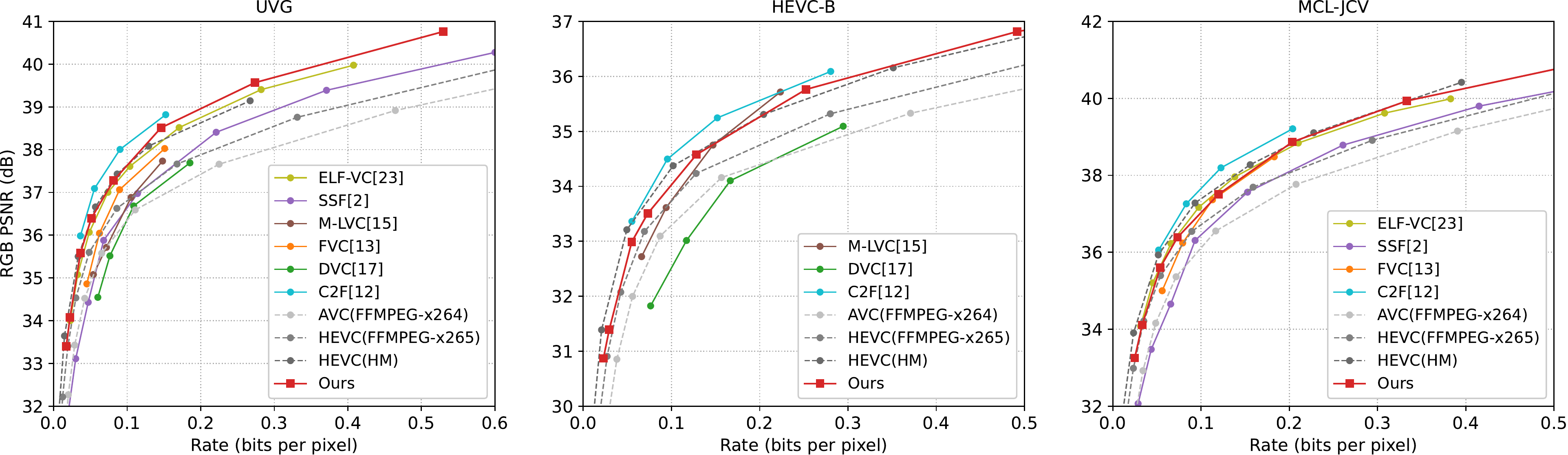}
    \end{subfigure}
    \vspace{1mm}
    \begin{subfigure}{1.0\textwidth}
      \raggedleft
      \includegraphics[width=.99\textwidth]{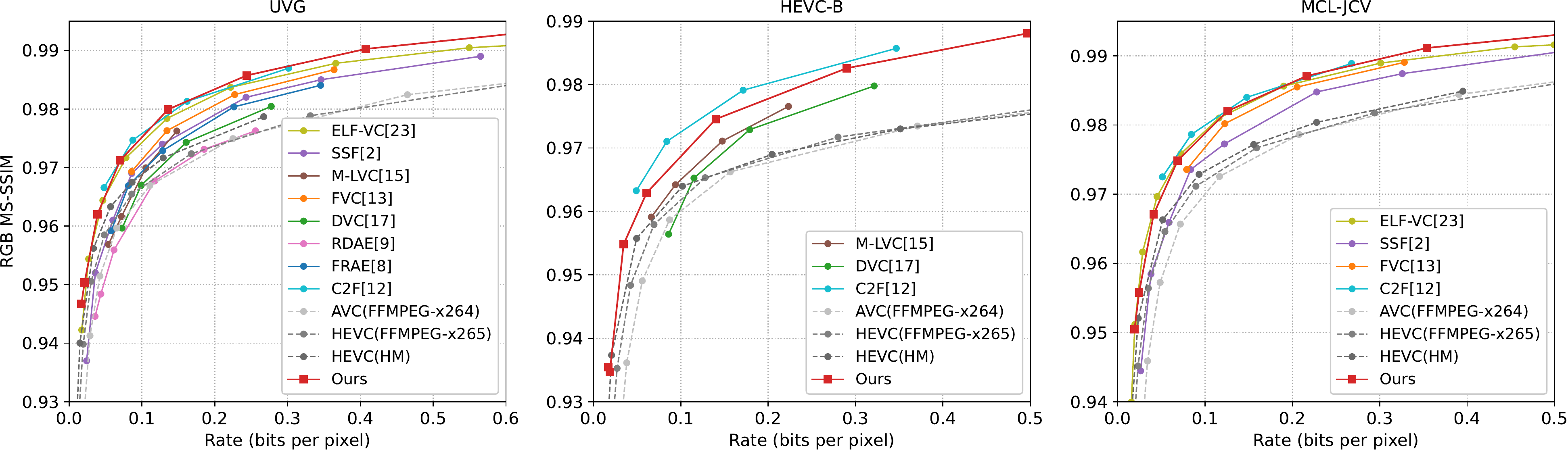}
    \end{subfigure}
    \vspace{-0.5em}
    \caption{RGB rate-distortion curves for the UVG, HEVC Class-B, and MCL-JCV datasets.
    }
    \label{fig:experiments:results:rgbrdcurves}
\end{figure*}

We measure rate-distortion performance on the UVG, MCL-JCV and HEVC-B datasets. 
We use $\psnr$ and $\msssim$ as distortion measures, and measure bitrate in bits per pixel.
We compare our RGB and YUV420 codecs to the reference implementation of HEVC called HM-16.25~\cite{hm} as well the the ffmpeg implementation of HEVC and AVC. 
We always use Group of Pictures size ``infinite``, even if some neural baselines use different GOP size.
HM is used in the low-delay-P configuration, and ffmpeg is used with all the default parameters but with B-frames disabled. 
Please see appendix~\ref{appendix:commands} for the used HM and ffmpeg commands. 
We compare our RGB codec with representative neural video coding methods including DVC~\cite{dvc}, SSF~\cite{ssf}, M-LVC~\cite{lin2020mlvc}, FVC~\cite{hu2021fvc}, ELF-VC~\cite{elfvc}, RDAE~\cite{habibian2019video}, FRAE~\cite{golinski2020frae}, and C2F~\cite{zhihao2022c2f}.
The RGB and YUV420 rate-distortions results are shown in Figures \ref{fig:experiments:results:rgbrdcurves} and \ref{fig:experiments:results:yuvrdcurves}, respectively. %In Figure~\ref{fig:experiments:results:yuvrdcurves},
Here, SSF-YUV is a re-implementation of SSF with the required YUV420 modifications.

In RGB colorspace, we outperform almost all of the compared methods in terms of both $\psnr$ and $\msssim$, except the very recent C2F~\cite{zhihao2022c2f}.
In YUV420 colorspace, classical codecs dominate on YUV $\psnr$, but our codec obtains much better $\msssim$ on all the datasets.

Many modern neural codecs use motion compensation and residual coding, and we emphasize that our method is often applicable to  those codecs as well.
Here, we applied our method to the SSF architecture and therefore measured the performance improvement 
with respect to a (reimplemented) SSF and SSF-YUV baseline in Tables \ref{tab:experiments:results:bdrates_rgb} and \ref{tab:experiments:results:bdrates_yuv}, respectively. 
We see that performance improvements are substantial, \eg, more than 30\% on the UVG dataset. 
Hence, although we are not generating state-of-the-art results in the RGB colorspace, our method has the potential to be added to a stronger base method and improve the state-of-the-art.

%%%%%%%%%%%%%%%%%%%%%%%%%%%%%%%%%%%%%%%%%%%%%%%%%%%%
\begin{figure*}
    \begin{subfigure}{1.0\textwidth}
      \raggedleft
      \includegraphics[width=.975\textwidth]{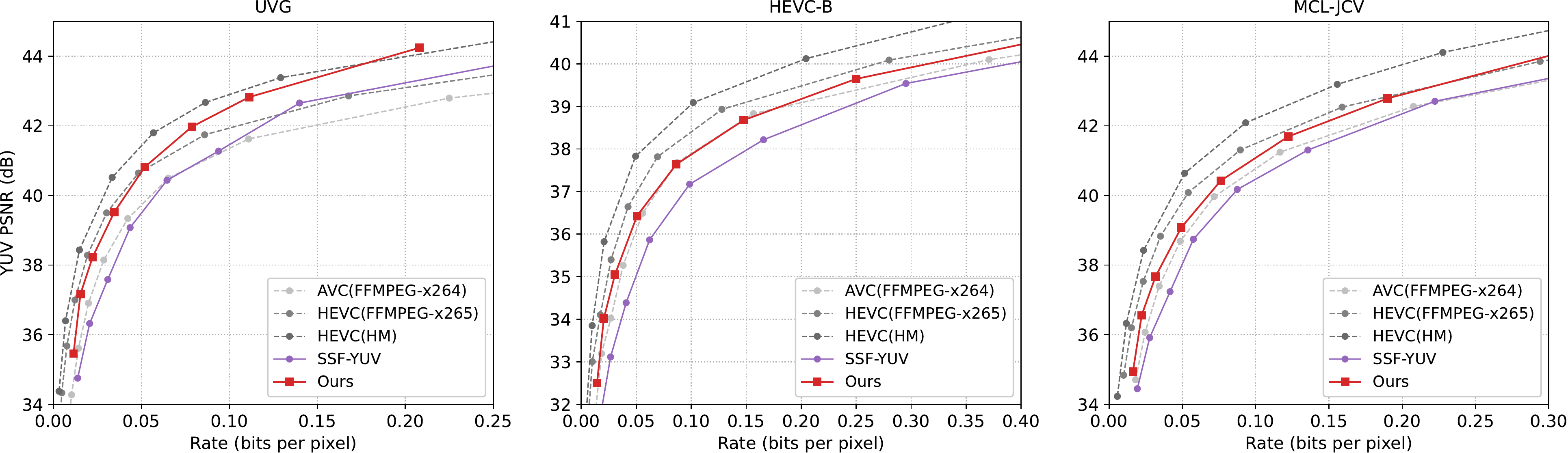}
      %\caption{}
      %\label{fig:experiments:results:rgbrdcurvesa}
    \end{subfigure}
    \begin{subfigure}{1.0\textwidth}
      \raggedleft
      \includegraphics[width=.99\textwidth]{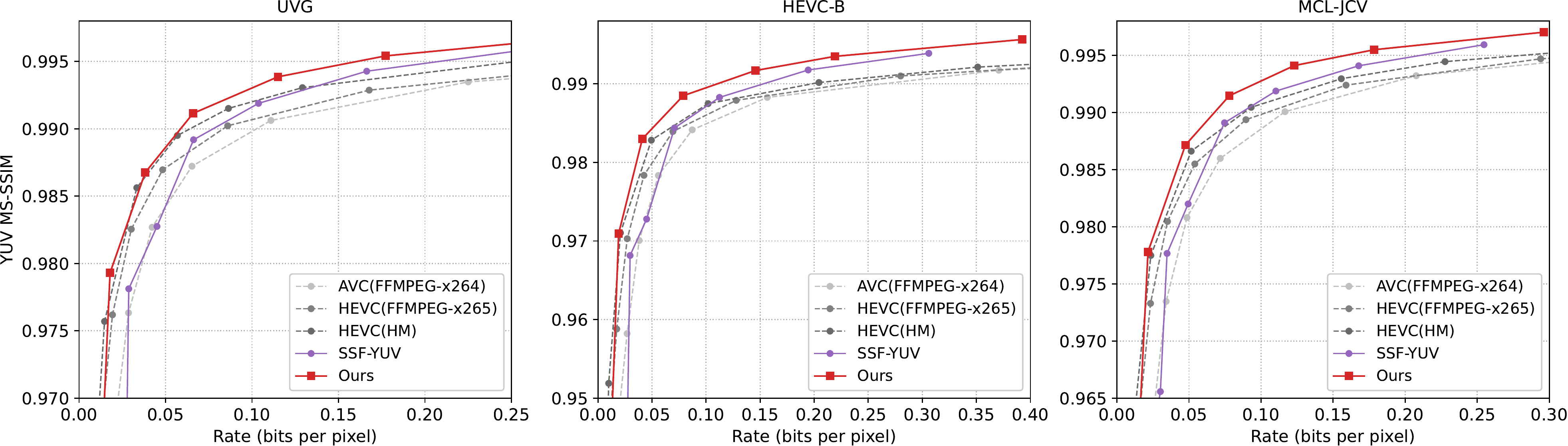}
      %\caption{}
      %\label{fig:experiments:results:rgbrdcurvesb}
    \end{subfigure}
    \vspace{-0.5em}
    \caption{YUV420 rate-distortion curves for the UVG, HEVC Class-B, and MCL-JCV datasets.}
    \label{fig:experiments:results:yuvrdcurves}
    %\vspace{-1.5em}
\end{figure*}

%-----------------------------------------
\subsection{Ablations}

To separate the effect of each component of our method, we perform ablation studies.
We incrementally add the flow predictor, residual predictor, and connections between the bottleneck of the predictors and autoencoders ($+$conditions).
The corresponding Bjontegaard-Delta~\cite{bjontegaard2001bdrate} (BD)-rate gains are shown in Table \ref{tab:experiments:ablations:bdrates}.
Each addition leads to strong BD-rate gains, and the flow predictor has the biggest effect for the high-framerate UVG dataset.

%%%%%%%%%%%%%%%%%%%%%%%%%%%%%%%%%%%%%%%%%
\begin{table}[t!]
    \centering
    \small
    \begin{tabular}{l rrr}
        \toprule
            Component       & UVG       & HEVC-B    & MCL-JCV   \\ 
        \midrule
            base model      & 0.0\%     & 0.0\%     & 0.0\%     \\
            +flow predictor & -29.60\%  & -9.24\%   & -9.88\%   \\
            +res predictor  & -33.73\%  & -20.37\%  & -10.34\%  \\
            +conditions     & -38.01\%  & -26.51\%  & -16.48\%  \\
        \bottomrule
    \end{tabular}
    \vspace{-0.5em}
    \caption{BD rate savings for RGB PSNR with respect to the SSF base model. Lower is better.}
    \label{tab:experiments:ablations:bdrates}
    \vspace{-1.5em}
\end{table}

Additionally, we measure bitrate savings for the flow and residual bitstreams individually to demonstrate that both $\flowae$ and $\resae$ benefit from our method. 
To measure BD-rate per component, we use the output PSNR as distortion and use the individual bitstream sizes per component as rate.
The BD-rate gains versus PSNR for the UVG, HEVC class B, and MCL-JCV datasets are shown in Figure~\ref{fig:experiments:results:bdrates_per_psnr}. 
Here, it can be observed that the predictors deliver significant rate savings on both flow and residual bitstreams across all PSNR values. 

For completeness, we list two training techniques that did not work for us in early experiments and were therefore abandoned.
Training with auxiliary losses for reconstructions resulting from flow and residual prediction, $\reconflowpredicted{t}$ and $\reconrespredicted{t}$, did not work better than using only a distortion loss defined on the reconstructions $\recon{t}$.
Temporal loss modulation as in~\cite{elfvc} resulted in $<1\%$ rate savings, so we omitted it for simplicity.

\begin{figure*}
    \centering
    \begin{subfigure}{.33\textwidth}
      \centering
      \includegraphics[width=.9\linewidth]{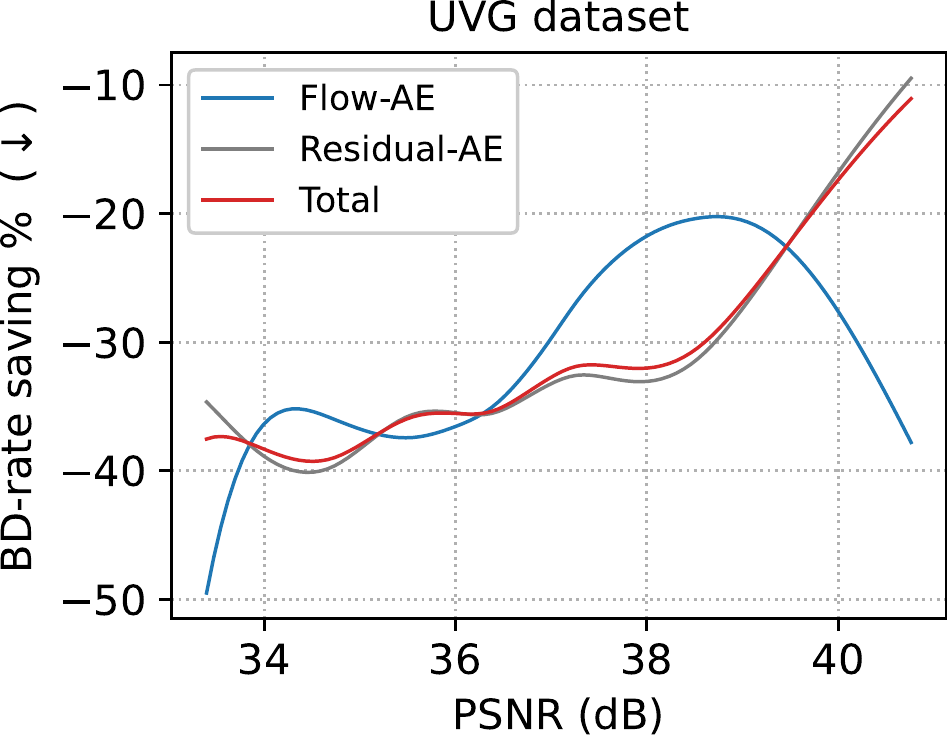}
    \end{subfigure}%
    \begin{subfigure}{.33\textwidth}
      \centering
      \includegraphics[width=.9\linewidth]{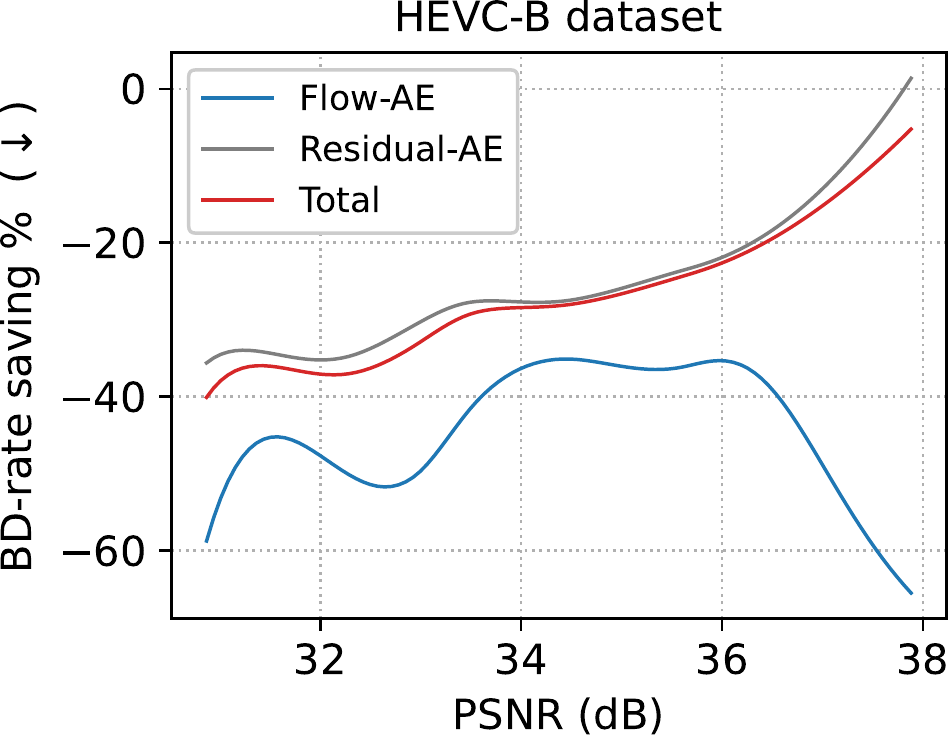}
    \end{subfigure}
    \begin{subfigure}{.33\textwidth}
      \centering
      \includegraphics[width=.9\linewidth]{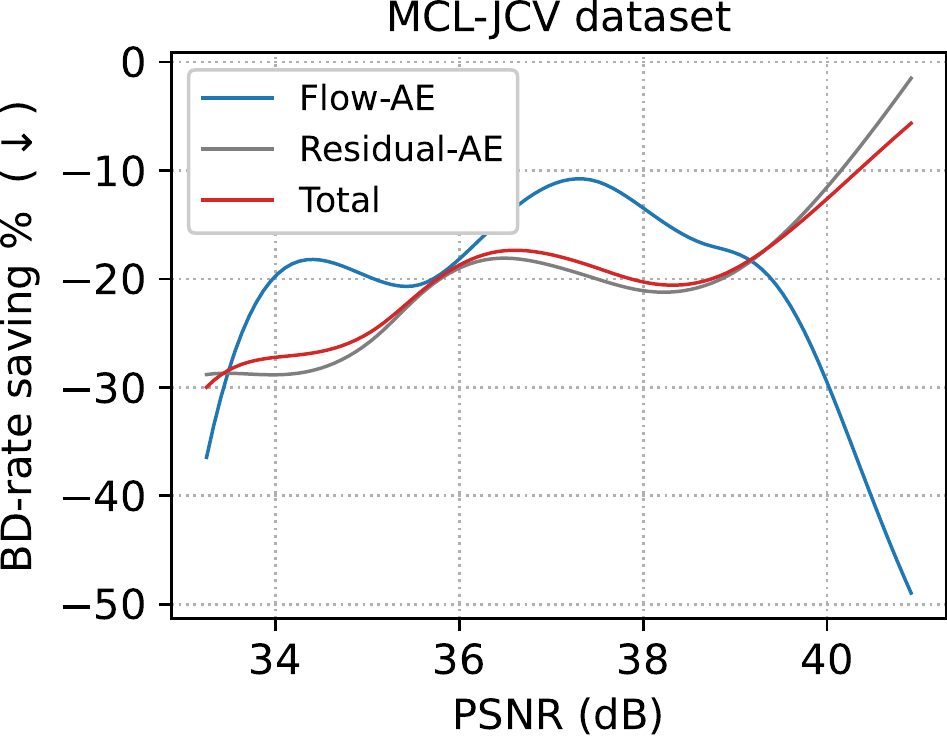}
    \end{subfigure}%
    \vspace{-0.5em}
    \caption{BD-rate savings versus RGB PSNR for flow and residual bitstreams and in-total for a SSF baseline. Lower is better.}
    \label{fig:experiments:results:bdrates_per_psnr}
    \vspace{-0.5em}
\end{figure*}

\section{Conclusion}

In this work, we demonstrate that improved use of receiver-side information substantially improves the rate-distortion performance of neural video codecs.
By having the decoder predict the flow and residual at a current timestep based on previously transmitted information, our codec only needs to transmit a correction, saving bits in the process.
This results in Bjontegaard-Delta rate savings of up to 30\% compared to a Scale-Space Flow baseline on common video benchmark datasets. 
The required architectural modification is straightforward, requires only lightweight components, and is applicable to other neural video codecs as well.

Furthermore, we train our model on both the RGB and YUV420 input domains with only minor architectural modifications, 
and show strong performance in both settings compared to neural and standard baselines.
As standard codecs were designed to optimize YUV420 PSNR, our codec does not outperform HM on this metric in the low bitrate regime, but substantially outperform the HM codec on YUV420 MS-SSIM.
Ablations show that both the flow and residual predictors contribute to the final performance.

\paragraph{Impact statement}

It is possible that learned codecs exacerbate biases present in the training data.
Nevertheless, we believe that improved video coding efficiency has a net positive impact on the world by reducing bandwidth and storage space needs, and by improving visual quality for video applications such as video conferencing.

%-------------------------------------------------------------------------

\clearpage

{\small
\bibliographystyle{ieee_fullname}
\bibliography{bibliography}
}

\clearpage

\newpage
%\appendix 

\section{Supplementary material}

%-------------------------
\subsection{Commands for standard codecs}
\label{appendix:commands}
We use the following commands to obtain compressed videos for standard codecs.
For ffmpeg, we use the medium preset, disable B-frames and otherwise use default settings.
For HM, we use HM-16.25 with the LowDelay-P config, which can be obtained from \href{https://vcgit.hhi.fraunhofer.de/jvet/HM/-/tags/HM-16.25}{https://vcgit.hhi.fraunhofer.de/jvet/HM/-/tags/HM-16.25}.

\begin{small}
\begin{lstlisting}
# For ffmpeg x264
ffmpeg -y -f rawvideo \ 
  -pix_fmt yuv420p \
  -s:v <width>x<height> \ 
  -r <framerate> \ 
  -i <input.yuv> \
  -c:v libx264 \
  -preset medium \  
  -crf <crf> \ 
  -x264-params bframes=0 \
  <output>

# For ffmpeg x265
ffmpeg -y -f rawvideo \ 
  -pix_fmt yuv420p \
  -s:v <width>x<height> \ 
  -r <framerate> \ 
  -i <input.yuv> \ 
  -c:v libx265 \
  -preset medium \  
  -crf <crf> \ 
  -x265-params bframes=0 \
  <output>

# For HM-16.25 LowDelayP
./bin/TAppEncoderStatic -c \ 
  ./cfg/encoder_lowdelay_P_main.cfg \
  -i <input.yuv> \
  --InputBitDepth=8 \ 
  -wdt <width> \
  -hgt <height> \
  -fr <framerate> \
  -f <numframes> \ 
  -q <qp> \
  -o <output>
\end{lstlisting}
\end{small}

%-------------------------
\subsection{Detailed computational complexity}
\label{appendix:commands}

We report the number of model parameters and multiply-and-accumulate (MAC) operations for the RGB and YUV codec in Table~\ref{tab:appendix:compute:complexity}, separating the numbers for the sender (``send'') and receiver (``recv'') side.
MACs are normalized with respect to the number of input pixels to enable comparisons across input resolutions and input modalities.
For each group of five rows, the shown percentage is computed with respect to that group.
All percentages are rounded, and some columns do not sum to 100\% as a result.

Note that in order to transmit data, the modules are run in their entirety due to the design of the SSF codec.
Compute for the sender side is therefore identical to the total compute as reported in Table~\ref{tab:experiments:compute:complexity} in the main text, and we repeat those numbers here for completeness.

On the receiver side, the predictors are run in their entirety, but only the decoder and hyper-decoder of each of the hyperprior modules has to be run.
We see that the computational complexity of our predictors is low on the receiver side still: 7.4\% of total parameters, 6.2\% of total MACs for the RGB receiver, and only 5.6\% for the YUV420 codec.

As a minor detail, for the SSF model, the residual decoder does not have to be run on the sender side for the very last frame in a Group of Pictures.
As this has a negligible effect on compute when coding with a large GoP size we do not take this into account here.

%%%%%%%%%%%%%%%%%%%%%%%%%%%%%%%%%%%%%%
\begin{table}[t!]
    \centering
    \small
\begin{tabular}{clrrrr}
\toprule
 & Module           & \multicolumn{2}{l}{Parameters} & \multicolumn{2}{l}{KMACs / pixel} \\ 
\midrule
\parbox[t]{2mm}{\multirow{5}{*}{\rotatebox[origin=c]{90}{RGB send}}}
& $\iframeae$    &   9.414M &    30.4\% &   198.2  &  30.6\% \\
& $\flowae$      &  10.038M &    32.4\% &   210.2  &  32.4\% \\
& $\resae$       &  10.028M &    32.5\% &   207.8  &  32.0\% \\
& $\flownet$     &   0.748M &     2.4\% &    16.2  &   2.5\% \\
& $\resnet$      &   0.748M &     2.4\% &    16.2  &   2.5\% \\
\midrule
\parbox[t]{2mm}{\multirow{5}{*}{\rotatebox[origin=c]{90}{RGB recv}}}
& $\iframeae$    &   5.790M &    28.8\% &   158.9 &   30.1\% \\    
& $\flowae$      &   6.410M &    31.9\% &   168.6 &   31.9\% \\    
& $\resae$       &   6.410M &    31.9\% &   168.5 &   31.9\% \\    
& $\flownet$     &   0.748M &     3.7\% &    16.2 &    3.1\% \\    
& $\resnet$      &   0.748M &     3.7\% &    16.2 &    3.1\% \\    
\midrule
\parbox[t]{2mm}{\multirow{5}{*}{\rotatebox[origin=c]{90}{YUV420 send}}}
& $\iframeae$    &    9.496M &   30.4\% &   208.0  &  30.5\% \\
& $\flowae$      &   10.124M &   32.4\% &   225.7  &  33.0\% \\
& $\resae$       &   10.111M &   32.3\% &   217.6  &  31.9\% \\
& $\flownet$     &    0.751M &    2.4\% &    16.3  &   2.4\% \\
& $\resnet$      &    0.750M &    2.4\% &    15.3  &   2.2\% \\
\midrule
\parbox[t]{2mm}{\multirow{5}{*}{\rotatebox[origin=c]{90}{YUV420 recv}}}
& $\iframeae$    &   5.880M &    28.9\% &   171.1 &   29.9\% \\
& $\flowae$      &   6.500M &    31.9\% &   188.8 &   33.0\% \\
& $\resae$       &   6.490M &    31.9\% &   180.7 &   31.6\% \\
& $\flownet$     &   0.751M &     3.7\% &    16.3 &    2.9\% \\
& $\resnet$      &   0.750M &     3.7\% &    15.3 &    2.7\% \\
\midrule
\bottomrule
\end{tabular} 
    \vspace{-0.5em}
    \caption{Parameters and MACs per pixel of our RGB and YUV420 codecs.}
    \label{tab:appendix:compute:complexity}
    \vspace{-1.5em}
\end{table}

\end{document}